\documentclass[a4paper,aps,twocolumn]{revtex4}
\RequirePackage[english]{babel}
\RequirePackage[latin1]{inputenc}
\RequirePackage[T1]{fontenc}
\RequirePackage{mathrsfs}
\RequirePackage{amsmath}
\RequirePackage{amssymb}
\RequirePackage{amsbsy}
\RequirePackage{bm}
\begin{document}
\title{\bf{Least-Order Torsion-Gravity for Fermion Fields, and the\\
Non-Linear Potentials in the Standard Models}}
\author{Luca Fabbri}
\affiliation{INFN \& Dipartimento di Fisica, Universit\`{a} di Bologna,\\
Via Irnerio 46, 40126 Bologna, ITALY}
\date{\today}
\begin{abstract}
We will consider the least-order torsional completion of gravity for a spacetime filled with fermionic Dirac matter fields, and we study the effects of the background-induced non-linear potentials for the matter field themselves in view of their effects for both standard models of physics: from the one of cosmology to that of particles, we will discuss the mechanisms of generation of the cosmological constant and particle masses as well as the phenomenology of leptonic weak-like forces and neutrino oscillations, the problem of zero-point energy, how there can be neutral massive fields as candidates for dark matter, and avoidance of gravitational singularity formation; we will show the way in which all these different effects can nevertheless be altogether described in terms of just a single model, which will be thoroughly discussed in the end.
\end{abstract}
\maketitle
\section*{Introduction}
In the present paper, we will consider the most general case of least-order derivative torsional completion of gravity describing spacetimes filled with $\frac{1}{2}$-spin fermionic Dirac matter fields: the requirement of least-order derivative means that the torsion-gravitational action will be constituted by quadratic torsion terms and it will be linear in the curvature, like for the usual Sciama-Kibble completion of Einstein gravity, and also Maxwell electrodynamics will be considered; the $\frac{1}{2}$-spin spinorial matter field may be introduced in terms of an action with only one derivative of the field itself, as usual. The total action will be then analyzed by obtaining the field equations, in which we will perform the customary treatment of separating torsion and having it substituted in terms of the spin of the spinorial matter, with the consequence that in the non-trivial background in which matter finds place there will be non-linear interactions of the matter fields.

Usually, these non-linearities are neglected, but in this paper, nevertheless, these non-linearities will not be neglected, and their effects on the dynamics of the matter field involved will be analyzed: we will see that there will be interesting consequences that will stretch from the standard model of cosmology to the standard model of particles, regarding the problem of the cosmological constant and its relations to the mechanism of spontaneous breaking of a given symmetry and the zero-point energy, the possibility to have massless neutrino oscillations, the existence of neutral massive fields as candidates for dark matter, and also the possibility to avoid the problem of singularity formation; from the perspective of contents, most of these results have already been established elsewhere in the literature, but always in models that were isolated from one another, missing any systematization.

The aim of this article is to show how the presented theory will be capable of addressing the effective weak-like forces among lepton fields, the oscillations between neutrinos, the dynamics of dark matter, as different effects of the same dynamics, that is how these effects can be altogether described in terms of a unique model.

Finally, we will see that the torsionally-induced non-linear potentials within the Dirac matter field equations and the fermionic-field quantization prescription have intrinsic similarities, which are discussed eventually.
\section{Least-Order Torsion-Gravity for Dirac Matter}
\subsection{Geometric and Kinematic Quantities}
To begin, we shall introduce the geometric and kinematic quantities we will employ, and although such introduction is quite well known it is nevertheless necessary in order to have the paper somewhat self-contained.

All along the present paper, the geometry we will employ is based on a $(1\!+\!3)$-dimensional spacetime and it will be a Riemann-Cartan geometry, that is described by the metric but also by the torsion: the process of raising/lowering indices in tensors is possible by introducing two tensors $g^{\mu\nu}$ and $g_{\mu\nu}$ which have to be considered as fundamental quantities, and if we require that raising and lowering the same index leaves the initial tensor unchanged then these two tensors have to be symmetric and one the inverse of the other, that is they must have all properties of a Riemannian metric, and so they will be called metric tensors; differential properties must preserve covariance, and thus they are given by covariant derivatives $D_{\mu}$ which can be defined after introducing the connection $\Gamma^{\alpha}_{\mu\nu}$ whose antisymmetric part in the two lower indices $\Gamma^{\alpha}_{\mu\nu}\!\!\!
-\!\!\Gamma^{\alpha}_{\nu\mu}\!\!=\!\!\! Q^{\alpha}_{\phantom{\alpha}\mu\nu}$ is nevertheless a tensor, which is called Cartan torsion tensor. Since we demand that the process of raising/lowering indices be possible also for tensors that are derivatives of some other tensor then the condition $D_{\mu}g_{\alpha\beta}\!=\!0$ must hold, and this condition means that the covariant differentiation does not affect the metric structure, and for this reason such condition is called metric-compatibility; in this paper the torsion tensor will be assumed to be completely antisymmetric, for the reason that follows: the metric-compatibility ensures local Lorentz structure to be preserved \cite{h,h-h-k-n}, and as a consequence we also have that the system of coordinates in which locally the metric is Minkowskian and the connection vanishes coincide, and on the other hand, a completely antisymmetric torsion ensures that there is a single symmetric part in the connection given by
\begin{eqnarray}
&\Gamma^{\alpha}_{\beta\mu}=\!=\!\frac{1}{2}g^{\alpha\rho}(\partial_{\beta}g_{\mu\rho}
\!+\!\partial_{\mu}g_{\beta\rho}\!-\!\partial_{\rho}g_{\mu\beta}\!+\!Q_{\rho\beta\mu})
\label{connectioninitial}
\end{eqnarray}
as the most general decomposition. So, the conditions of metric-compatibility and completely antisymmetric torsion together ensure that we can always find a coordinate system where locally both the metric is Minkowskian and the symmetric part of the connection vanishes: this fact allows the implementation of the light-cone structure and the local free-fall as discussed in 
\cite{m-l,a-l}. Therefore, causality and the principle of equivalence can be mathematically implemented \cite{xy}, and gravitation is geometrized.

From the metric we can define the completely antisymmetric tensor of Levi-Civita $\varepsilon_{\rho\mu\nu\alpha}$ as usual; we also introduce the covariant derivative $\nabla_{\mu}$ defined in terms of the connection $\Lambda^{\alpha}_{\mu\nu}$ that is called Levi-Civita connection, which is the simplest connection in the sense that it is symmetric in the two lower indices and it is written in terms of the metric alone. We have that the metric-compatibility $\nabla_{\mu}g_{\alpha\beta}\!=\!0$ holds identically; the completely antisymmetric torsion can be written according to the following expression $Q_{\alpha\mu\nu}\!=\! \varepsilon_{\alpha\mu\nu\sigma} W^{\sigma}$ in terms of this tensor as the dual of the axial vector $W_{\alpha}$ which encodes all information about the torsion, but it may be much simpler to employ in some situations: then we have that the covariant derivatives of the Levi-Civita tensor vanish identically as a consequence of the metric-compatibility conditions above, and the above connection is written as
\begin{eqnarray}
&\Gamma^{\alpha}_{\beta\mu}\!=\!\Lambda^{\alpha}_{\beta\mu}
\!+\!\frac{1}{2}g^{\alpha\rho}\varepsilon_{\rho\beta\mu\sigma}W^{\sigma}
\label{connection}
\end{eqnarray}
as an equivalent decomposition, so that the most general connection can be decomposed into the simplest Levi-Civita connection plus axial vector torsion contributions.

This formalism (with Greek indices) is called coordinate formalism, and it is possible to introduce an equivalent but different formalism (with Latin indices) called Lorentz formalism, with the advantage that in it the most general coordinate transformation is converted without loss of generality into the special Lorentz transformation, whose specific form may be explicited and therefore given in terms of other, different representations: in Lorentz formalism, the metric is decomposed according to the following $g^{\alpha\nu}\!=\!e^{\alpha}_{p}e^{\nu}_{i} \eta^{pi}$ and $g_{\alpha\nu}\!=\!e_{\alpha}^{p}e_{\nu}^{i} \eta_{pi}$ in terms of the tetrad basis given by $e_{\alpha}^{i}$ and the dual $e^{\alpha}_{i}$ and in terms of the constant metric $\eta_{ij}$ defined to have Minkowskian structure, the one which will have to be preserved by the Lorentz transformation; differential properties preserving also this type of covariance are defined analogously in terms of the covariant derivative $D_{\mu}$ defined after introducing the spin-connection $\omega^{i}_{\phantom{i}j\nu}$ from which we can define no torsion at all. However, we have that metric-compatibilities $D_{\mu}e_{i}^{\nu}\!=\!0$ and $D_{\mu}\eta_{ij}\!=\!0$ hold; even if we may define no torsion from the spin-connection, it is however possible by employing such metric-compatibilities to convert the above torsion into this formalism as
\begin{eqnarray}
&-Q^{k}_{\alpha\rho}=\partial_{\alpha}e_{\rho}^{k}-\partial_{\rho}e_{\alpha}^{k}
+\omega^{k}_{\phantom{i}p\alpha}e_{\rho}^{p}-\omega^{k}_{\phantom{i}p\rho}e_{\alpha}^{p}
\label{translation}
\end{eqnarray} 
as it is easy to check directly: we notice that the metric-compatibility conditions applied on the Minkowskian metric implies that $\omega^{ip}_{\phantom{ip}\alpha} \!=\! -\omega^{pi}_{\phantom{pi}\alpha}$ spelling the antisymmetry in the two Lorentz indices of the spin-connection itself, and ensuring local Lorentz structure to be preserved as we will see, while the metric-compatibility applied on the tetrad implies that the above connection can be transformed into the spin-connection according to
\begin{eqnarray}
&\omega^{i}_{\phantom{i}p\alpha}\!=\!
e^{i}_{\sigma}e^{\rho}_{p}(e^{k}_{\rho}\partial_{\alpha}e^{\sigma}_{k}
\!+\!\Gamma^{\sigma}_{\rho\alpha})
\label{spin-connection}
\end{eqnarray} 
in the most general case possible. Thus these two metric-compatibility conditions are what ensures that the coordinate formalism and the Lorentz formalism are equivalent, and that the latter is better equipped to incorporate local Lorentz structures: this will be important when dealing with spinors. As anticipated, with the Lorentz formalism we can look for different representations.

Because one possible different representation is the complex representation, it is useful to introduce also the geometry of complex fields, in which the transformation is given by a complex unitary phase: gauge covariant derivatives $D_{\mu}$ are defined after introducing the gauge connection $A_{\nu}$ as it is known. We will next see why it is important to introduce such abelian gauge structure.

We will introduce the Lorentz group in complex representations, and we will restrict ourselves to the least-spin given by the $\frac{1}{2}$-spin representation: such a representation can be achieved through the introduction of the $\boldsymbol{\gamma}_{a}$ matrices such that $\{\boldsymbol{\gamma}_{a},\boldsymbol{\gamma}_{b}\}\!=\!2\boldsymbol{\mathbb{I}}\eta_{ab}$ from which one may define the matrices $\boldsymbol{\sigma}_{ab}\!=\!\frac{1}{4}
[\boldsymbol{\gamma}_{a},\boldsymbol{\gamma}_{b}]$ as the infinitesimal generators of the Lorentz transformation that is written in the complex $\frac{1}{2}$-spin representation, and these matrices also verify the relations $\{\boldsymbol{\gamma}_{a},\boldsymbol{\sigma}_{bc}\}\!=\!
i\varepsilon_{abcd} \boldsymbol{\pi}\boldsymbol{\gamma}^{d}$ implicitly defining the $\boldsymbol{\pi}$ matrix that will be used to define the left-handed and the right-handed projectors; differential properties are given by the most general spinorial covariant derivative $\boldsymbol{D}_{\mu}$ defined upon introduction of the most general spinorial connection $\boldsymbol{\Omega}_{\mu}$ and there is no torsion defined in its terms. Conditions $\boldsymbol{D}_{\mu}\boldsymbol{\gamma}_{a}\!=\!0$ are valid automatically as identities; notice that in this case it is not even possible to express the already known torsion tensor in spinorial form: from the conditions of metric-compatibility it is possible to see that the spinorial connection can be written according to the form
\begin{eqnarray}
&\boldsymbol{\Omega}_{\rho}
=\frac{1}{2}\omega^{ij}_{\phantom{ij}\rho}\boldsymbol{\sigma}_{ij}
\!+\!iqA_{\rho}\mathbb{I}
\label{spinorial-connection}
\end{eqnarray}
as a general decomposition. Thus we finally see that the antisymmetry in the two Lorentz indices of the spin-connection is mirrored by the antisymmetry in the two indices of the generators of the Lorentz transformation, and this is what ensures local Lorentz structure to be preserved in this formalism: the most general spinorial connection is not however exhausted by the Lorentz group since there is still room for an abelian field which may now be identified with the abelian gauge field above, and so the above is actually the most general decomposition of the spinorial connection. This is the connection for a spinor field of $q$ charge $\frac{1}{2}$-spin and where there is no appearance of the mass of the spinor for the moment.

So far we have introduced a very compact overview of the general formalism we will employ in the present paper dealing with first-order derivatives, but of course it is also possible to go at the second order to see what additional structures may come out: the very first of these, and possibly the most fundamental, is the curvature tensor
\begin{eqnarray}
&G^{\rho}_{\phantom{\rho}\xi\mu\nu}
=\partial_{\mu}\Gamma^{\rho}_{\xi\nu}-\partial_{\nu}\Gamma^{\rho}_{\xi\mu}
+\Gamma^{\rho}_{\sigma\mu}\Gamma^{\sigma}_{\xi\nu}
-\Gamma^{\rho}_{\sigma\nu}\Gamma^{\sigma}_{\xi\mu}
\end{eqnarray}
antisymmetric in both first and second pair of indices and such that it verifies the cyclic permutation property
\begin{eqnarray}
\nonumber
&D_{\kappa}Q^{\rho}_{\phantom{\rho}\mu \nu}
\!+\!D_{\nu}Q^{\rho}_{\phantom{\rho} \kappa \mu}
\!+\!D_{\mu}Q^{\rho}_{\phantom{\rho} \nu \kappa}\!+\!\\
\nonumber
&+Q^{\pi}_{\phantom{\pi} \nu \kappa}Q^{\rho}_{\phantom{\rho}\mu \pi}
\!+Q^{\pi}_{\phantom{\pi}\mu \nu}Q^{\rho}_{\phantom{\rho}\kappa \pi}
\!+Q^{\pi}_{\phantom{\pi}\kappa \mu}Q^{\rho}_{\phantom{\rho}\nu \pi}-\\
&-G^{\rho}_{\phantom{\rho}\kappa \nu \mu}
\!-\!G^{\rho}_{\phantom{\rho}\mu \kappa \nu}
\!-\!G^{\rho}_{\phantom{\rho}\nu \mu \kappa}\equiv0
\label{JBtorsion}
\end{eqnarray}
and because of these antisymmetry properties, the curvature has a single contraction given by $G^{\rho}_{\phantom{\rho}\mu\rho\nu}\!=\!G_{\mu\nu}$ with contraction $G_{\eta\nu}g^{\eta\nu}\!=\!G$ called Ricci tensor and scalar respectively; the curvature comes along with its torsionless counterpart given in terms of the analogous expression
\begin{eqnarray}
&R^{\rho}_{\phantom{\rho}\xi\mu\nu}
=\partial_{\mu}\Lambda^{\rho}_{\xi\nu}-\partial_{\nu}\Lambda^{\rho}_{\xi\mu}
+\Lambda^{\rho}_{\sigma\mu}\Lambda^{\sigma}_{\xi\nu}
-\Lambda^{\rho}_{\sigma\nu}\Lambda^{\sigma}_{\xi\mu}
\label{metriccurvature}
\end{eqnarray}
antisymmetric in both first and second pair of indices, and such that it verifies the cyclic permutation property given by $R^{\rho}_{\phantom{\rho}\kappa \nu \mu}\!+\!
R^{\rho}_{\phantom{\rho}\mu \kappa \nu}\!+\!R^{\rho}_{\phantom{\rho}\nu \mu \kappa}\equiv0$ as it is again easy to check from the above identities, and in the same way we have a single contraction $R^{\rho}_{\phantom{\rho}\mu\rho\nu}\!=\!R_{\mu\nu}$ with contraction given by $R_{\eta\nu}g^{\eta\nu}\!=\!R$ as above: we then have that
\begin{eqnarray}
\nonumber
&G^{\rho}_{\phantom{\rho}\xi\mu\nu}=R^{\rho}_{\phantom{\rho}\xi\mu\nu}
+\frac{1}{2}(\nabla_{\mu}Q^{\rho}_{\phantom{\rho}\xi\nu}
\!-\!\nabla_{\nu}Q^{\rho}_{\phantom{\rho}\xi\mu})+\\
&+\frac{1}{4}(Q^{\rho}_{\phantom{\rho}\sigma\mu}Q^{\sigma}_{\phantom{\sigma}\xi\nu}
\!-\!Q^{\rho}_{\phantom{\rho}\sigma\nu}Q^{\sigma}_{\phantom{\sigma}\xi\mu})
\label{separation}
\end{eqnarray}
is their most general decomposition. In the equivalent Lorentz formalism the curvature is written according to
\begin{eqnarray}
&G^{i}_{\phantom{i}j\mu\nu}
=\partial_{\mu}\omega^{i}_{j\nu}-\partial_{\nu}\omega^{i}_{j\mu}
+\omega^{i}_{p\mu}\omega^{p}_{j\nu}-\omega^{i}_{p\nu}\omega^{p}_{j\mu}
\end{eqnarray}
again antisymmetric in both the coordinate and the Lorentz indices: we have that we may write
\begin{eqnarray}
&G^{i}_{\phantom{i}j\mu\nu}\!=\!G^{\rho}_{\phantom{\rho}\sigma\mu\nu} e^{\sigma}_{j}e^{i}_{\rho}
\end{eqnarray}
as it should be for consistency, since such an object is a tensor and therefore the passage from general coordinate to Lorentz formalisms must be given in terms of a simple index renaming. The gauge connection has an analogous curvature that is given by the similar expression
\begin{eqnarray}
&F_{\alpha\beta}=\partial_{\alpha}A_{\beta}-\partial_{\beta}A_{\alpha}
\end{eqnarray}
which is antisymmetric in its indices. Some additional identities for these curvatures are given by the following
\begin{eqnarray}
\nonumber
&D_{\mu}G^{\nu}_{\phantom{\nu}\iota \kappa \rho}
\!+\!D_{\kappa}G^{\nu}_{\phantom{\nu}\iota \rho \mu}
\!+\!D_{\rho}G^{\nu}_{\phantom{\nu}\iota \mu \kappa}+\\
&+G^{\nu}_{\phantom{\nu}\iota \beta \mu}Q^{\beta}_{\phantom{\beta}\rho \kappa}
\!+\!G^{\nu}_{\phantom{\nu}\iota \beta \kappa}Q^{\beta}_{\phantom{\beta}\mu \rho}
\!+\!G^{\nu}_{\phantom{\nu}\iota \beta \rho}Q^{\beta}_{\phantom{\beta}\kappa \mu}\equiv0
\label{JBcurvature}
\end{eqnarray}
again as it is easy to check directly: in Lorentz formalism this identity remains unchanged, as expected; also
\begin{eqnarray}
&\partial_{\nu}F_{\alpha\sigma}\!+\!\partial_{\sigma}F_{\nu\alpha}
\!+\!\partial_{\alpha}F_{\sigma\nu}=0
\end{eqnarray}
for the abelian gauge field. The importance of the curvature tensor comes from the fact that with it it becomes possible to express the commutator of covariant derivatives as $[D_{\mu},D_{\nu}]V^{\alpha}\!\!=\! Q^{\rho}_{\phantom{\rho}\mu\nu}D_{\rho}V^{\alpha}\!+\!
G^{\alpha}_{\phantom{\alpha}\rho\mu\nu}V^{\rho}$ in the case of vectors, and similarly with one curvature term for each tensorial index in the case of generic tensors of any order.

In terms of these geometrical quantities it is possible to build the spinorial equivalent of the curvature tensor
\begin{eqnarray}
&\boldsymbol{G}_{\mu\nu}
\!=\!\partial_{\mu}\boldsymbol{\Omega}_{\nu}\!-\!\partial_{\nu}\boldsymbol{\Omega}_{\mu}
\!+\!\boldsymbol{\Omega}_{\mu}\boldsymbol{\Omega}_{\nu}
\!-\!\boldsymbol{\Omega}_{\nu}\boldsymbol{\Omega}_{\mu}
\end{eqnarray}
antisymmetric in its two indices: it can be written as
\begin{eqnarray}
&\boldsymbol{G}_{\mu\nu}=\frac{1}{2}G^{ij}_{\phantom{ij}\mu\nu}\boldsymbol{\sigma}_{ij}
\!+\!iqF_{\mu\nu}\mathbb{I}
\label{spinorial-tensor}
\end{eqnarray} 
in its most general decomposition. In terms of this curvature the commutator of spinorial covariant derivatives has the form $[\boldsymbol{D}_{\mu},\boldsymbol{D}_{\nu}]\psi\!=\! Q^{\rho}_{\phantom{\rho}\mu\nu} \boldsymbol{D}_{\rho}\psi\!+\!\boldsymbol{G}_{\mu\nu}\psi$ clearly showing that the gravitational and the electrodynamic curvatures may be formally united together within the most general curvature tensor that can be defined to act upon the spinor fields we are taking into consideration.
\subsection{Dynamical Equations}
Having defined the kinematic quantities we will employ, we proceed to have these kinematic quantities coupled together by constructing the dynamical field equations, which will be taken at the least-order derivative.

A first way we have to construct the most general system of field equations is to start from geometrical identities, postulating the geometric field equations that will convert these geometric identities into conservation laws, and postulating matter field equations that ensure those conservation laws be verified: this geometric construction at the least-order derivative possible is easy, and so starting from the Jacobi-Bianchi geometrical identities for the torsion tensor in their fully contracted form
\begin{eqnarray}
&D_{\rho}Q^{\rho\mu\nu}\!-\!G^{\mu\nu}\!+\!G^{\nu\mu}\!\equiv\!0
\end{eqnarray}
together with the Jacobi-Bianchi geometrical identities for the curvature tensor in their fully contracted form
\begin{eqnarray}
&D_{\mu}(G^{\mu\rho}\!-\!\frac{1}{2}Gg^{\mu\rho})\!-\!G_{\mu\beta}Q^{\beta\mu\rho}
\!-\!\frac{1}{2}G^{\mu\eta\beta\rho}Q_{\beta\eta\mu}\!\equiv\!0
\end{eqnarray}
and geometrical identities obtained from the commutator of covariant derivatives applied to the case of the gauge strength in their fully contracted form given by
\begin{eqnarray}
&D_{\rho}(D_{\nu}F^{\nu\rho}\!+\!\frac{1}{2}Q^{\rho\mu\nu}F_{\mu\nu})\!=\!0
\end{eqnarray}
it is possible to see that the least-order field equations, describing the coupling at the least-order derivative possible, are given for the coupling between the completely antisymmetric torsion and spin in a purely algebraical form according to the following relationship
\begin{eqnarray}
&Q^{\rho\mu\nu}=-kS^{\rho\mu\nu}
\label{torsion}
\end{eqnarray}
together with the field equations describing the coupling between the non-symmetric curvature and energy as
\begin{eqnarray}
\nonumber
&\left(\frac{1-k}{2k}\right)(D_{\mu}Q^{\mu\rho\alpha}
\!-\!\frac{1}{2}Q^{\theta\sigma\rho}Q_{\theta\sigma}^{\phantom{\theta\sigma}\alpha}
\!+\!\frac{1}{4}Q^{\theta\sigma\pi}Q_{\theta\sigma\pi}g^{\rho\alpha})+\\
\nonumber
&+(G^{\rho\alpha}\!-\!\frac{1}{2}Gg^{\rho\alpha}\!-\!\Lambda g^{\rho\alpha})-\\
&-\frac{1}{2}(\frac{1}{4}g^{\rho\alpha}F^{2}
\!-\!F^{\rho\theta}F^{\alpha}_{\phantom{\alpha}\theta})
=\frac{1}{2}T^{\rho\alpha}
\label{curvature}
\end{eqnarray}
and the field equations describing the coupling between the derivative of the gauge strength and the current
\begin{eqnarray}
&\frac{1}{2}F_{\mu\nu}Q^{\rho\mu\nu}\!+\!D_{\sigma}F^{\sigma\rho}=J^{\rho}
\label{gauge}
\end{eqnarray}
and these convert the above identities into conservation laws for the completely antisymmetric spin tensor as
\begin{eqnarray}
&D_{\rho}S^{\rho\mu\nu}\!+\!\frac{1}{2}(T^{\mu\nu}\!-\!T^{\nu\mu})=0
\label{conservationspin}
\end{eqnarray}
together with the conservation law for the non-symmetric energy tensor given according to the following form
\begin{eqnarray}
&D_{\mu}T^{\mu\rho}\!-\!T_{\mu\sigma}Q^{\sigma\mu\rho}\!+\!S_{\beta\mu\sigma}G^{\sigma\mu\beta\rho}
\!+\!J_{\beta}F^{\beta\rho}=0
\label{conservationenergy}
\end{eqnarray}
and the conservation law for the current vector
\begin{eqnarray}
D_{\rho}J^{\rho}=0
\label{conservationcurrent}
\end{eqnarray}
which will have to be verified once the matter field equations are satisfied; then, it is possible to see that these conservation laws are verified by the conserved quantities given by the completely antisymmetric spin tensor
\begin{eqnarray}
\nonumber
&S^{\rho\mu\nu}=a_{1}\frac{i}{4}
\overline{\psi}_{1}\{\boldsymbol{\gamma}^{\rho}\!,\!\boldsymbol{\sigma}^{\mu\nu}\}\psi_{1}+\\
&+a_{2}\frac{i}{4}
\overline{\psi}_{2}\{\boldsymbol{\gamma}^{\rho}\!,\!\boldsymbol{\sigma}^{\mu\nu}\}\psi_{2}
\label{spin}
\end{eqnarray}
with non-symmetric energy tensor in the form
\begin{eqnarray}
\nonumber
&T^{\rho\alpha}=\frac{i}{2}(\overline{\psi}_{1}\boldsymbol{\gamma}^{\rho}\!\boldsymbol{D}^{\alpha}\psi_{1}
\!-\!\boldsymbol{D}^{\alpha}\overline{\psi}_{1}\!\boldsymbol{\gamma}^{\rho}\psi_{1})-\\
\nonumber
&-(a_{1}\!-\!1)D_{\mu}(\frac{i}{4}\overline{\psi}_{1}\{\boldsymbol{\gamma}^{\mu}
\!,\!\boldsymbol{\sigma}^{\rho\alpha}\}\psi_{1})+\\
\nonumber
&+(a_{1}\!-\!1)\frac{i}{4}\overline{\psi}_{1}\{\boldsymbol{\gamma}^{\rho}
\!,\!\boldsymbol{\sigma}_{\mu\nu}\}\psi_{1}Q^{\alpha\mu\nu}-\\
\nonumber
&-\frac{1}{2}(a_{1}\!-\!1)\frac{i}{4}\overline{\psi}_{1}\{\boldsymbol{\gamma}^{\alpha}
\!,\!\boldsymbol{\sigma}_{\mu\nu}\}\psi_{1}Q^{\rho\mu\nu}+\\
\nonumber
&+\frac{i}{2}(\overline{\psi}_{2}\boldsymbol{\gamma}^{\rho}\!\boldsymbol{D}^{\alpha}\psi_{2}
\!-\!\boldsymbol{D}^{\alpha}\overline{\psi}_{2}\!\boldsymbol{\gamma}^{\rho}\psi_{2})-\\
\nonumber
&-(a_{2}\!-\!1)D_{\mu}(\frac{i}{4}\overline{\psi}_{2}\{\boldsymbol{\gamma}^{\mu}
\!,\!\boldsymbol{\sigma}^{\rho\alpha}\}\psi_{2})+\\
\nonumber
&+(a_{2}\!-\!1)\frac{i}{4}\overline{\psi}_{2}\{\boldsymbol{\gamma}^{\rho}
\!,\!\boldsymbol{\sigma}_{\mu\nu}\}\psi_{2}Q^{\alpha\mu\nu}-\\
&-\frac{1}{2}(a_{2}\!-\!1)\frac{i}{4}\overline{\psi}_{2}\{\boldsymbol{\gamma}^{\alpha}
\!,\!\boldsymbol{\sigma}_{\mu\nu}\}\psi_{2}Q^{\rho\mu\nu}
\label{energy}
\end{eqnarray}
and the current tensor in terms of the expression
\begin{eqnarray}
&J^{\rho}=q_{1}\overline{\psi}_{1}\boldsymbol{\gamma}^{\rho}\psi_{1}
\!+\!q_{2}\overline{\psi}_{2}\boldsymbol{\gamma}^{\rho}\psi_{2}
\label{current}
\end{eqnarray}
whenever the set of least-order field equations for the fermion fields that are given according to the expressions
\begin{eqnarray}
&i\boldsymbol{\gamma}^{\mu}\!\boldsymbol{D}_{\mu}\psi_{1}
\!+\!\frac{i}{8}(a_{1}\!-\!1)Q_{\rho\mu\nu}\boldsymbol{\gamma}^{\rho}\boldsymbol{\gamma}^{\mu}
\boldsymbol{\gamma}^{\nu}\psi_{1}\!-\!m_{1}\psi_{1}\!=\!0
\label{fermion1}\\
&i\boldsymbol{\gamma}^{\mu}\!\boldsymbol{D}_{\mu}\psi_{2}
\!+\!\frac{i}{8}(a_{2}\!-\!1)Q_{\rho\mu\nu}\boldsymbol{\gamma}^{\rho}\boldsymbol{\gamma}^{\mu}
\boldsymbol{\gamma}^{\nu}\psi_{2}\!-\!m_{2}\psi_{2}\!=\!0
\label{fermion2}
\end{eqnarray}
are satisfied as conditions on the fermionic fields.

Before proceeding, we notice that because in this geometric construction we start from geometric identities and from them we obtain the conservation laws after integration, and similarly the matter field equations are obtained from these conservation laws after integration, then the additional $\Lambda$, $m_{1}$ and $m_{2}$ constants can be interpreted as constants of integration, which can be added into the theory with the same meaning with which constants of integrations are taken into account, that is in order to maintain the highest degree of generality.

This geometric construction follows the spirit of deriving from geometry all physical equations, but it is also possible to follow a variational method in terms of which we can write the most general least-order derivative Lagrangian, and then vary it to obtain the corresponding least-order derivative field equations: the most general torsional-gravitational Lagrangian consists of both parity-odd as well as parity-even terms \cite{Hojman:1980kv}, although in the case of completely antisymmetric torsion it happens that the parity-odd terms are either vanishing as for the torsion contribution or surface terms as for the gravitational Holst term, and similarly for the gauge sector we may add both parity-odd and parity-even terms, although also in this case the parity-odd contributions due to the gauge strength are surface terms; the most general material Lagrangian for the fermion fields can be build again with both parity-odd and parity-even fermionic terms as done in \cite{Alexandrov:2008iy}, although it is also possible to demonstrate by employing Fierz rearrangements that all parity-odd contributions disappear. As a consequence, we may summarize this discussion by considering the most general Lagrangian as the one given in \cite{Fabbri:2013gza} with the additional assumption of parity-conservation, or in an equivalent way as the one given in \cite{Fabbri:2012yg} with the additional hypothesis of having an axial vector torsion, with quadratic torsion contributions beside the gravitational contributions described by the usual Einstein term and electrodynamic contribution described by the usual Maxwell term, together with the Dirac term; additionally, we will include all interacting terms between all the fields, which in this specific case reduce to a supplementary term coupling torsion to the spinors. The most general Lagrangian for the case of two spinor fields is given according to
\begin{eqnarray}
\nonumber
&L\!=\!(\frac{k-1}{4k})Q_{\alpha\nu\sigma}Q^{\alpha\nu\sigma}\!+\!G\!+\!2\Lambda
\!+\!\frac{1}{4}F^{\alpha\nu}F_{\alpha\nu}-\\
\nonumber
&-\frac{i}{2}(\overline{\psi}_{1}\boldsymbol{\gamma}^{\mu}\!\boldsymbol{D}_{\mu}\psi_{1}
\!-\!\boldsymbol{D}_{\mu}\overline{\psi}_{1}\!\boldsymbol{\gamma}^{\mu}\psi_{1})-\\
\nonumber
&-\frac{1}{8}(a_{1}\!-\!1)i\overline{\psi}_{1}\boldsymbol{\gamma}^{\nu} \boldsymbol{\gamma}^{\sigma}\boldsymbol{\gamma}^{\pi}\psi_{1}Q_{\nu\sigma\pi}
\!+\!m_{1}\overline{\psi}_{1}\psi_{1}-\\
\nonumber
&-\frac{i}{2}(\overline{\psi}_{2}\boldsymbol{\gamma}^{\mu}\!\boldsymbol{D}_{\mu}\psi_{2}
\!-\!\boldsymbol{D}_{\mu}\overline{\psi}_{2}\!\boldsymbol{\gamma}^{\mu}\psi_{2})-\\
&-\frac{1}{8}(a_{2}\!-\!1)i\overline{\psi}_{2}\boldsymbol{\gamma}^{\nu} \boldsymbol{\gamma}^{\sigma}\boldsymbol{\gamma}^{\pi}\psi_{2}Q_{\nu\sigma\pi}
\!+\!m_{2}\overline{\psi}_{2}\psi_{2}
\label{actionleast}
\end{eqnarray}
without terms mixing the spinorial fields with one another and up to dimension four, as it can be checked.

With this Lagrangian we may perform its variation with respect to all the independents fields involved thus obtaining the corresponding field equations, starting from the completely antisymmetric torsion-spin coupling field equations that are given in the following form
\begin{eqnarray}
\nonumber
&Q^{\rho\mu\nu}\!=\!-k(a_{1}\frac{i}{4}
\overline{\psi}_{1}\{\boldsymbol{\gamma}^{\rho}\!,\!\boldsymbol{\sigma}^{\mu\nu}\}\psi_{1}+\\
&+a_{2}\frac{i}{4}
\overline{\psi}_{2}\{\boldsymbol{\gamma}^{\rho}\!,\!\boldsymbol{\sigma}^{\mu\nu}\}\psi_{2})
\label{torsion-spin}
\end{eqnarray}
which come together with the non-symmetric curvature-energy coupling field equations given according to
\begin{eqnarray}
\nonumber
&\left(\frac{1-k}{2k}\right)(D_{\mu}Q^{\mu\rho\alpha}
\!-\!\frac{1}{2}Q^{\theta\sigma\rho}Q_{\theta\sigma}^{\phantom{\theta\sigma}\alpha}
\!+\!\frac{1}{4}Q^{\theta\sigma\pi}Q_{\theta\sigma\pi}g^{\rho\alpha})+\\
\nonumber
&+(G^{\rho\alpha}\!-\!\frac{1}{2}Gg^{\rho\alpha}\!-\!\Lambda g^{\rho\alpha})-\\
\nonumber
&-\frac{1}{2}(\frac{1}{4}g^{\rho\alpha}F^{2}
\!-\!F^{\rho\theta}F^{\alpha}_{\phantom{\alpha}\theta})=\\
\nonumber
&=\frac{i}{4}(\overline{\psi}_{1}\boldsymbol{\gamma}^{\rho}\!\boldsymbol{D}^{\alpha}\psi_{1}
\!-\!\boldsymbol{D}^{\alpha}\overline{\psi}_{1}\!\boldsymbol{\gamma}^{\rho}\psi_{1})-\\
\nonumber
&-\frac{1}{2}(a_{1}\!-\!1)D_{\mu}(\frac{i}{4}\overline{\psi}_{1}\{\boldsymbol{\gamma}^{\mu}
\!,\!\boldsymbol{\sigma}^{\rho\alpha}\}\psi_{1})+\\
\nonumber
&+\frac{1}{2}(a_{1}\!-\!1)\frac{i}{4}\overline{\psi}_{1}\{\boldsymbol{\gamma}^{\rho}
\!,\!\boldsymbol{\sigma}_{\mu\nu}\}\psi_{1}Q^{\alpha\mu\nu}-\\
\nonumber
&-\frac{1}{4}(a_{1}\!-\!1)\frac{i}{4}\overline{\psi}_{1}\{\boldsymbol{\gamma}^{\alpha}
\!,\!\boldsymbol{\sigma}_{\mu\nu}\}\psi_{1}Q^{\rho\mu\nu}+\\
\nonumber
&+\frac{i}{4}(\overline{\psi}_{2}\boldsymbol{\gamma}^{\rho}\!\boldsymbol{D}^{\alpha}\psi_{2}
\!-\!\boldsymbol{D}^{\alpha}\overline{\psi}_{2}\!\boldsymbol{\gamma}^{\rho}\psi_{2})-\\
\nonumber
&-\frac{1}{2}(a_{2}\!-\!1)D_{\mu}(\frac{i}{4}\overline{\psi}_{2}\{\boldsymbol{\gamma}^{\mu}
\!,\!\boldsymbol{\sigma}^{\rho\alpha}\}\psi_{2})+\\
\nonumber
&+\frac{1}{2}(a_{2}\!-\!1)\frac{i}{4}\overline{\psi}_{2}\{\boldsymbol{\gamma}^{\rho}
\!,\!\boldsymbol{\sigma}_{\mu\nu}\}\psi_{2}Q^{\alpha\mu\nu}-\\
&-\frac{1}{4}(a_{2}\!-\!1)\frac{i}{4}\overline{\psi}_{2}\{\boldsymbol{\gamma}^{\alpha}
\!,\!\boldsymbol{\sigma}_{\mu\nu}\}\psi_{2}Q^{\rho\mu\nu}
\label{curvature-energy}
\end{eqnarray}
and with the gauge-current coupling field equations as
\begin{eqnarray}
&\frac{1}{2}F_{\mu\nu}Q^{\rho\mu\nu}\!+\!D_{\sigma}F^{\sigma\rho}
\!=\!q_{1}\overline{\psi}_{1}\boldsymbol{\gamma}^{\rho}\psi_{1}
\!+\!q_{2}\overline{\psi}_{2}\boldsymbol{\gamma}^{\rho}\psi_{2}
\label{gauge-current}
\end{eqnarray}
complemented by the fermionic field equations
\begin{eqnarray}
&i\boldsymbol{\gamma}^{\mu}\!\boldsymbol{D}_{\mu}\psi_{1}
\!+\!\frac{i}{8}(a_{1}\!-\!1)Q_{\rho\mu\nu}\boldsymbol{\gamma}^{\rho}\boldsymbol{\gamma}^{\mu}
\boldsymbol{\gamma}^{\nu}\psi_{1}\!-\!m_{1}\psi_{1}\!=\!0
\label{fermionic1}\\
&i\boldsymbol{\gamma}^{\mu}\!\boldsymbol{D}_{\mu}\psi_{2}
\!+\!\frac{i}{8}(a_{2}\!-\!1)Q_{\rho\mu\nu}\boldsymbol{\gamma}^{\rho}\boldsymbol{\gamma}^{\mu}
\boldsymbol{\gamma}^{\nu}\psi_{2}\!-\!m_{2}\psi_{2}\!=\!0
\label{fermionic2}
\end{eqnarray}
as the most general system of field equations, and coinciding with the system of field equations found above.

Notice that the torsional constant $k$ is not the gravitational constant, which has been set to the unity, while the gauge coupling constants are given by the charges as defined above, and the constants $a_{1}$ and $a_{2}$ are entirely new constants interpreted as the two coupling constants with which torsion couples to the two fermionic fields respectively, according to the two additional terms that have been included in the Lagrangian for generality.

This action can also be written in the form in which all curvatures and derivatives are decomposed in terms of their torsionless counterparts plus torsional contributions and since the torsion-spin coupling field equations are algebraic they can be used to have torsion substituted in terms of the spin of the fermionic fields yielding
\begin{eqnarray}
\nonumber
&L=R\!+\!2\Lambda\!+\!\frac{1}{4}F^{\mu\nu}F_{\mu\nu}-\\
\nonumber
&-\frac{i}{2}(\overline{\psi}_{1}\boldsymbol{\gamma}^{\mu}\boldsymbol{\nabla}_{\mu}\psi_{1}
\!-\!\boldsymbol{\nabla}_{\mu}\overline{\psi}_{1}\boldsymbol{\gamma}^{\mu}\psi_{1})-\\
\nonumber
&-\frac{1}{2}X_{12}\overline{\psi}_{1}\boldsymbol{\gamma}_{\mu}\boldsymbol{\pi}\psi_{1}
\overline{\psi}_{2}\boldsymbol{\gamma}^{\mu}\boldsymbol{\pi}\psi_{2}-\\
\nonumber
&-\frac{1}{2}Y_{1}\overline{\psi}_{1}\boldsymbol{\gamma}_{\mu}\boldsymbol{\pi}\psi_{1}
\overline{\psi}_{1}\boldsymbol{\gamma}^{\mu}\boldsymbol{\pi}\psi_{1}
\!+\!m_{1}\overline{\psi}_{1}\psi_{1}-\\
\nonumber
&-\frac{i}{2}(\overline{\psi}_{2}\boldsymbol{\gamma}^{\mu}\boldsymbol{\nabla}_{\mu}\psi_{2}
\!-\!\boldsymbol{\nabla}_{\mu}\overline{\psi}_{2}\boldsymbol{\gamma}^{\mu}\psi_{2})-\\
\nonumber
&-\frac{1}{2}X_{12}\overline{\psi}_{1}\boldsymbol{\gamma}_{\mu}\boldsymbol{\pi}\psi_{1}
\overline{\psi}_{2}\boldsymbol{\gamma}^{\mu}\boldsymbol{\pi}\psi_{2}-\\
&-\frac{1}{2}Y_{2}\overline{\psi}_{2}\boldsymbol{\gamma}_{\mu}\boldsymbol{\pi}\psi_{2}
\overline{\psi}_{2}\boldsymbol{\gamma}^{\mu}\boldsymbol{\pi}\psi_{2}
\!+\!m_{2}\overline{\psi}_{2}\psi_{2}
\label{actionleastdecomposed}
\end{eqnarray}
which will eventually yield the system of the field equations already in the form in which torsion has been replaced with spin-spin contact fermionic interactions.

Or alternatively, these field equations can be obtained from the previous field equations after decomposing all curvatures and derivatives in the corresponding torsionless curvatures and derivatives plus torsional contributions written as the spin of fermionic fields, yielding the symmetric curvature-energy coupling field equations as
\begin{eqnarray}
\nonumber
&(R^{\rho\alpha}\!-\!\frac{1}{2}Rg^{\rho\alpha}\!-\!\Lambda g^{\rho\alpha})-\\
\nonumber
&-\frac{1}{2}(\frac{1}{4}g^{\rho\alpha}F^{2}
\!-\!F^{\rho\theta}F^{\alpha}_{\phantom{\alpha}\theta})=\\
\nonumber
&=\frac{i}{8}(\overline{\psi}_{1}\boldsymbol{\gamma}^{\rho}\boldsymbol{\nabla}^{\alpha}\psi_{1}
\!-\!\boldsymbol{\nabla}^{\alpha}\overline{\psi}_{1}\boldsymbol{\gamma}^{\rho}\psi_{1}+\\
\nonumber
&+\overline{\psi}_{1}\boldsymbol{\gamma}^{\alpha}\boldsymbol{\nabla}^{\rho}\psi_{1}
\!-\!\boldsymbol{\nabla}^{\rho}\overline{\psi}_{1}\boldsymbol{\gamma}^{\alpha}\psi_{1})+\\
\nonumber
&+\frac{i}{8}(\overline{\psi}_{2}\boldsymbol{\gamma}^{\rho}\boldsymbol{\nabla}^{\alpha}\psi_{2}
\!-\!\boldsymbol{\nabla}^{\alpha}\overline{\psi}_{2}\boldsymbol{\gamma}^{\rho}\psi_{2}+\\
\nonumber
&+\overline{\psi}_{2}\boldsymbol{\gamma}^{\alpha}\boldsymbol{\nabla}^{\rho}\psi_{2}
\!-\!\boldsymbol{\nabla}^{\rho}\overline{\psi}_{2}\boldsymbol{\gamma}^{\alpha}\psi_{2})+\\
\nonumber
&+\frac{1}{4}X_{12}\overline{\psi}_{1}\boldsymbol{\gamma}_{\mu}\boldsymbol{\pi}\psi_{1}
\overline{\psi}_{2}\boldsymbol{\gamma}^{\mu}\boldsymbol{\pi}\psi_{2}g^{\alpha\rho}+\\
\nonumber
&+\frac{1}{4}X_{12}\overline{\psi}_{2}\boldsymbol{\gamma}^{\mu}\boldsymbol{\pi}\psi_{2}
\overline{\psi}_{1}\boldsymbol{\gamma}_{\mu}\boldsymbol{\pi}\psi_{1}g^{\alpha\rho}+\\
\nonumber
&+\frac{1}{4}Y_{1}\overline{\psi}_{1}\boldsymbol{\gamma}_{\mu}\boldsymbol{\pi}\psi_{1}
\overline{\psi}_{1}\boldsymbol{\gamma}^{\mu}\boldsymbol{\pi}\psi_{1}g^{\alpha\rho}+\\
&+\frac{1}{4}Y_{2}\overline{\psi}_{2}\boldsymbol{\gamma}_{\mu}\boldsymbol{\pi}\psi_{2}
\overline{\psi}_{2}\boldsymbol{\gamma}^{\mu}\boldsymbol{\pi}\psi_{2}g^{\alpha\rho}
\label{gravitational}
\end{eqnarray}
and with the gauge-current coupling field equations
\begin{eqnarray}
&\nabla_{\sigma}F^{\sigma\rho}\!=\!q_{1}\overline{\psi}_{1}\boldsymbol{\gamma}^{\rho}\psi_{1}
\!+\!q_{2}\overline{\psi}_{2}\boldsymbol{\gamma}^{\rho}\psi_{2}
\label{electrodynamical}
\end{eqnarray}
together with the fermionic field equations
\begin{eqnarray}
\nonumber
&i\boldsymbol{\gamma}^{\mu}\boldsymbol{\nabla}_{\mu}\psi_{1}
\!+\!X_{12}\overline{\psi}_{2}\boldsymbol{\gamma}_{\rho}\boldsymbol{\pi}\psi_{2}
\boldsymbol{\gamma}^{\rho}\boldsymbol{\pi}\psi_{1}+\\
&+Y_{1}\overline{\psi}_{1}\boldsymbol{\gamma}_{\rho}\boldsymbol{\pi}\psi_{1}
\boldsymbol{\gamma}^{\rho}\boldsymbol{\pi}\psi_{1}\!-\!m_{1}\psi_{1}\!=\!0
\label{fermionical1}\\
\nonumber
&i\boldsymbol{\gamma}^{\mu}\boldsymbol{\nabla}_{\mu}\psi_{2}
\!+\!X_{12}\overline{\psi}_{1}\boldsymbol{\gamma}_{\rho}\boldsymbol{\pi}\psi_{1}
\boldsymbol{\gamma}^{\rho}\boldsymbol{\pi}\psi_{2}+\\
&+Y_{2}\overline{\psi}_{2}\boldsymbol{\gamma}_{\rho}\boldsymbol{\pi}\psi_{2}
\boldsymbol{\gamma}^{\rho}\boldsymbol{\pi}\psi_{2}\!-\!m_{2}\psi_{2}\!=\!0
\label{fermionical2}
\end{eqnarray}
showing that the system of field equations has reduced to the one we would have had without torsion but supplemented with non-linear potentials of a specific structure.

Notice that in these expressions we have renamed the coupling constant $3ka_{1}a_{2}\!=\!16X_{12}\!\equiv\!16X_{21}$ with the self-coupling constants $3ka_{1}^{2}\!=\!16Y_{1}$ and $3ka_{2}^{2}\!=\!16Y_{2}$ showing that the constants originally related to torsion are now re-expressed as constants describing the coupling of the spinorial fields with one another and each of them with itself, as if there were no torsion but we could perform on every fermion field an independent renormalization.
\section{Non-Linear Potential Effects}
\subsection{Problems of the Standard Models}
So far, we have seen what are the field equations we will employ, and in particular that the matter field equations are endowed with non-linear potentials: in the matter field equations these potentials closely resemble those that are usually taken into account for bosonization and the later condensation in condensed-state physics, with the difference that now they occur in general, and in the following of the paper we will take these non-linearities to their consequences; we will see that there are intriguing effects that arise in conjunction to the open problems of the standard models of cosmology and particle physics.

The list of open problems in the standard models of cosmology and particle physics starts with a problem that is common to both, that is the problem of the Cosmological Constant: quantum field theory would appear to be a very good theory in terms of the quantity of predictions that have been experimentally confirmed if it were not for the fact that among all such predictions is one which is the worst ever made: the value of the cosmological constant it predicts is one-hundred and twenty orders of magnitude off the empirical one; this happens because the cosmological constant that can theoretically be expected may be present as a generic universal constant, with no known value, but it might also receive further contributions coming from symmetry breaking, and its subsequent mechanism of mass generation, and the zero-point energy, as the minimal energy of vacuum fluctuations, both very large. As we have already said, the contribution that can always be present as a generic integration constant is by construction unknown, but even if we could invoke some principle like conformal symmetry to get rid of this unknown contribution, so to deal only with known contributions, we still have the problem that all these contributions are so large, and in principle unrelated, that a fine-tuning of about one-hundred and twenty digits is necessary to require they have an almost exact cancellation; to worsen things is the fact that the most relevant of such contributions is negative, while the cosmological constant is positive. This embarrassing situation has no solution that is entirely successful for now.

Of course, we should be open-minded enough to admit that some contributions may well be missing, but in this case there would simply be no way in which at the moment we could account for all possible contributions, and whether or not fine-tuning may be found, there would be no way to even start solving this problem; so, we might allow ourselves a little confidence, or hope, assuming that all contributions we have are actually all there is, but these contributions are one unknown and two known but large and with no definite sign, so any combination that aims to get any precise, small and positive value of the total cosmological constant would involve a flabbergasting amount of fine-tuning: unless this fine-tuning is found, there really seems to be no way out apart from the most dramatic of them all, and that is imagining that no such contributions are actually there in the first place.

Or better, that there can only be the bare cosmological constant, whose value will be the one chosen in nature, and that no further contributions can be present, so to circumvent any issue related to the values we have calculated they ought to have: because the contribution coming from the mechanism of mass generation is ultimately induced by a breakdown of some symmetry and the minimal energy of vacuum fluctuations is ultimately induced by the zero-point energy, the most straightforward solution is to insist on having no breaking of any symmetry nor zero-point energy at all. As already stated, these are quite drastic measures, but our aim is precisely that of showing that even these extreme cases can have reasonable solutions, so let us proceed in their investigation.

Dropping the mechanism of symmetry breaking requires two adjustments: one is that there must be an alternative mechanism of assigning masses to all the massive fields; the other is that there has to be a way to produce the observed phenomenology. The problem of finding an alternative mechanism needed to assign masses to massive fields may be solved by acknowledging that we do not actually need an alternative mechanism of mass generation since bare masses may always be present, with values that will be those chosen in nature, since there is no theoretical reason why elementary fields should not have fundamental masses instead of generated ones, and for that matter the specific values of the fundamental masses are not less arbitrary than the Yukawa constants for the generated ones; the fact that now the situation is always asymmetric renders it clear that the phenomenology must be given in terms of interactions mediated by something different. In this case the non-linearities of the model may turn out to be just what we need.

However preposterous this idea might seem, interactions mediated by something else than gauge fields were already described as in the Nambu-Jona--Lasinio model presented in references 
\cite{n-j--l/1,n-j--l/2} and further works on the subject such as for instance \cite{g-n}, with mediators thought as composite states of fermions bound together by non-linear non-symmetric effective interactions: the idea is to recover the construction built in the references above but in the case of the leptonic fields \cite{s-g}, characterized by a non-chiral structure since electrons are massive and double-handed while neutrinos are supposed to be massless and single-handed: field equations (\ref{fermionical1}-\ref{fermionical2}) become
\begin{eqnarray}
&i\boldsymbol{\gamma}^{\mu}\boldsymbol{\nabla}_{\mu}e
\!-\!X\overline{\nu}\boldsymbol{\gamma}_{\rho}\nu\boldsymbol{\gamma}^{\rho}\boldsymbol{\pi}e
\!-\!Y\overline{e}\boldsymbol{\gamma}_{\rho}e\boldsymbol{\gamma}^{\rho}e\!-\!me=0\\
&i\boldsymbol{\gamma}^{\mu}\boldsymbol{\nabla}_{\mu}\nu
\!-\!X\overline{e}\boldsymbol{\gamma}_{\rho}\boldsymbol{\pi}e\boldsymbol{\gamma}^{\rho}\nu=0
\end{eqnarray}
which can be Fierz rearranged into the equivalent
\begin{eqnarray}
\nonumber
&i\boldsymbol{\gamma}^{\mu}\boldsymbol{\nabla}_{\mu}e
\!+\!q\tan{\theta}Z_{\mu}\boldsymbol{\gamma}^{\mu}e
\!-\!\frac{g}{2\cos{\theta}}Z_{\mu}\boldsymbol{\gamma}^{\mu}e_{L}+\\
&+\frac{g}{\sqrt{2}}W_{\mu}^{\ast}\boldsymbol{\gamma}^{\mu}\nu
\!-\!He\!-\!iA\boldsymbol{\pi}e\!-\!me=0
\label{electron}\\
&i\boldsymbol{\gamma}^{\mu}\boldsymbol{\nabla}_{\mu}\nu
\!+\!\frac{g}{2\cos{\theta}}Z_{\mu}\boldsymbol{\gamma}^{\mu}\nu
\!+\!\frac{g}{\sqrt{2}}W_{\mu}\boldsymbol{\gamma}^{\mu}e_{L}=0
\label{neutrino}
\end{eqnarray}
so soon as we define the following composite states
\begin{eqnarray}
&Z_{\mu}\!=\!\frac{2X\cos{\theta}}{g|\sin{\theta}|^{2}}
\left[\frac{1}{2}(\overline{e}_{L}\boldsymbol{\gamma}_{\mu}e_{L}
\!-\!\overline{\nu}\boldsymbol{\gamma}_{\mu}\nu)
\!-\!|\sin{\theta}|^{2}\overline{e}\boldsymbol{\gamma}_{\mu}e\right]
\label{neutral}\\
&W_{\mu}\!=\!\left[\frac{\sqrt{2}X(1\!-\!4|\sin{\theta}|^{2})}{2g|\sin{\theta}|^{2}}\right]
\overline{e}_{L}\boldsymbol{\gamma}_{\mu}\nu
\label{charged}\\
&H\!=\!\left[Y\!+\!X(1\!-\!2|\sin{\theta}|^{2})\right]\overline{e}e\\
&A\!=\!\left[Y\!+\!X(1\!-\!2|\sin{\theta}|^{2})\right]i\overline{e}\boldsymbol{\pi}e
\end{eqnarray}
which in the present interpretation have to be thought as leptonic condensed states, those which give rise to the vector mediators of the weak interactions and to two additional scalar bosons, as it has been discussed in \cite{Fabbri:2010ux}.

In the present notation, we have suppressed the indices because the first spinor has been identified with the electron and the second spinor has been identified with the neutrino, and similarly we have written the torsional constants $X_{12}\!=\!X_{21}\!=\!X$ and $Y_{1}\!=\!Y$ re-expressing them in terms of the usual constants of the standard model.

According to expressions (\ref{neutral}-\ref{charged}) we have the following
\begin{eqnarray}
&\nabla_{\mu}Z^{\mu}\!=\!-\frac{X\cos{\theta}}{g|\sin{\theta}|^{2}}m\omega
\label{conservedneutral}\\
&\nabla_{\mu}W^{\mu}\!+\!iq\tan{\theta}Z^{\mu}W_{\mu}
\!=\!-\frac{\sqrt{2}X(1-4|\sin{\theta}|^{2})}{2g|\sin{\theta}|^{2}}m\rho
\label{conservedcharged}
\end{eqnarray}
in terms of the pseudo-scalar complex fields
\begin{eqnarray}
&\omega\!=\!i\overline{e}\boldsymbol{\pi}e\\
&\rho\!=\!i\overline{e}\boldsymbol{\pi}\nu\left[1\!+\!
\frac{Y\!-\!X(1\!+\!2|\sin{\theta}|^{2})}{Y\!+\!X(1\!-\!2|\sin{\theta}|^{2})}
\frac{H\!-\!iA}{m}\right]
\end{eqnarray}
whose form shows that the vector mediators are massive.

Incidentally, we notice that the partially conserved axial current of the charged vector loses all higher-order non-linear terms if $Y\!\!=\!\!X(1\!+\!2|\sin{\theta}|^{2})$ is assumed.

The attentive reader may have noticed that here it is assumed that neutrinos are in fact massless and single-handed while the problem of neutrino oscillation appears to require neutrino masses to work; because apart from this single circumstance neutrino masses do not seem to be expected in any other context, it may be wise to look for mechanisms of neutrino oscillation even if neutrinos are massless, as discussed in \cite{ds-g}: in that paper neutrinos are still double-handed although the right-handed neutrino is undetected yet, and so it may be wise to look for a mechanism of neutrino oscillation even if neutrinos are massless and single-handed, and to do that we consider again the field equations (\ref{fermionical1}-\ref{fermionical2}) adapted for this case as
\begin{eqnarray}
&i\boldsymbol{\gamma}^{\mu}\boldsymbol{\nabla}_{\mu}\nu_{1}
\!+\!\Xi\ \overline{\nu}_{2}\boldsymbol{\gamma}_{\rho}\nu_{2}\boldsymbol{\gamma}^{\rho}\nu_{1}=0\\
&i\boldsymbol{\gamma}^{\mu}\boldsymbol{\nabla}_{\mu}\nu_{2}
\!+\!\Xi\ \overline{\nu}_{1}\boldsymbol{\gamma}_{\rho}\nu_{1}\boldsymbol{\gamma}^{\rho}\nu_{2}=0
\end{eqnarray}
which can be written in a compact way for the doublet of neutrino fields $\overline{\nu}
\!=\!(\overline{\nu}_{1},\overline{\nu}_{2})$ in the equivalent form
\begin{eqnarray}
&i\boldsymbol{\gamma}^{\mu}(\boldsymbol{\nabla}_{\mu}\nu
\!+\!i\Xi\vec{A}_{\mu}\!\cdot\!\vec{\frac{\boldsymbol{\sigma}}{2}}\nu)=0
\end{eqnarray}
upon definition of the triplet of vectors
\begin{eqnarray}
&\vec{A}_{\mu}\!=\!\frac{2}{3}\overline{\nu}\boldsymbol{\gamma}_{\mu}\vec{\boldsymbol{\sigma}}\nu
\end{eqnarray}
showing that the mixing is formally a flavour-changing neutral current, even if the two neutrinos are single-handed and precisely because they are massless \cite{Fabbri:2010hz}.

In the present notation, there is a single torsional coupling constant $X_{12}\!=\!X_{21}\!=\!\Xi$ which can be re-expressed in terms of the oscillation length of the neutrinos.

In these two cases we have reached the point in which the non-linear effective interactions described above have come across some open problems in cosmology, but for the specific problem of the cosmological constant this construction only deals with the part concerning non-symmetric effective interactions, and we still need to address the part regarding the concept of zero-point energy.

Dropping the concept of zero-point energy requires one adjustment: there must be a way to recover the phenomenology of the Casimir effect. Or equivalently, it has to be possible to calculate the magnitude of the Casimir force without any reference to the zero-point energy.

Once again, this may seem preposterous, but such computation has already been done by exploiting alternatives sources as described in \cite{s}, and recently in \cite{j}.

The problem of the zero-point energy is very delicate because ultimately it touches the foundations of the theory of field quantization: the zero-point energy is the result of having fermion fields quantized with canonical anticommutation relations; getting rid of the zero-point energy means getting rid of the unity term in the anticommutation relations. Such a solution might look radical, but it merely means to place in normal-ordering all creation/annihilation operators, as it is commonly done in quantum field theories \cite{p-s}: we will not enter here into the details of the apparent contradiction that arises from the fact that in order for field quantization to work it seems necessary to abandon what constitutes the very essence of the quantization of fields, and we will simply remark that such a normal-ordering is equivalent to ask that there be no quantization of fields for fermions, apart from their being Grassmann-valued. Thus we may ask whether there are reasons for relativistic classical fields to necessitate adjustments in terms of quantization.

Focusing on fermionic fields, there is a problem that might be solved with field-quantization: we do know that fermionic fields do not undergo to the Fermi-Dirac statistic, nor have positive-defined energies or correct charge-conjugation transformations unless they are quantized with anticommutation relations; but once again there is no need to go so far as requiring fermionic anticommutation, because it is enough to require fermions to have Grassmann values. However, in the most extreme situation, we may renounce to Grassmann variables too.

Dropping the assumption of having fermionic fields to be Grassmann-valued means that we have to address all these problems: we have to reproduce the Fermi-Dirac statistic by finding a way in which two identical fermions cannot superpose, to ensure that all fermions have energies that are positive and to ensure a correct definition of matter and antimatter. Here again there is an alternative way of accomplishing all this, and again this way consists in using the non-linearities of the Dirac field equation.

First of all, the problem of finding a way in which identical fermions cannot superpose in the case of non-linear Dirac field equations is trivial, since in general this cannot happen: let $\psi$ and $\psi'$ be two fermions solutions of the Dirac equations $D[\psi]\!=\!0$ and $D[\psi']\!=\!0$ it is clear that the superposition $(\psi\!+\!\psi')$ of the two fermions cannot be solution of the Dirac equation $D[(\psi\!+\!\psi')]\!=\!0$ in general, and in particular if $\psi\!=\!\psi'$ the two fermions are identical and the Dirac equation $D[2\psi]\!\neq\!D[\psi]\!=\!0$ because of the non-linear term; so we may reverse the question by asking if there are situations in which such a superposition of fermions can still be solution of the Dirac equation despite the non-linearity. It is in fact possible for these two fermions to superpose as solutions of the Dirac equation when the non-linear term of their sum vanishes, and this is precisely what happens whenever the two fermions have opposite helicity states \cite{s-s}; thus the non-linear potential producing the effective repulsion that keeps apart two fermions when they are identical will nevertheless allow the superposition of two fermions with opposite helicity, entailing a dynamical version of the Pauli principle \cite{Fabbri:2010rw}. For bosons there is no non-linear potential creating an effective repulsion and the superposition of bosons is always possible. That bosons be described by tensorial fields that do not feel torsion while fermions are described by spinorial fields that do feel torsion is what discriminates them in the present way of dealing with the statistics. And bosons also have no problems of having an energy that is negative as much as fermions have, so again the solution may well be connected to the torsion tensor: as a matter of fact, the conservation laws of energy and spin (\ref{conservationspin}-\ref{conservationenergy}) make it clear that any fermion that suffers the reversal of the sign of its energy must also suffer the reversal of the sign of its spin, which means the inversion of the sign of torsion and thus the flip of the sign of the non-linear term within the Dirac field equations (\ref{fermion1}-\ref{fermion2}); thus fermions with negative energy are described by spinors that are not solution of the non-linear Dirac equation. The absence of the negative-energy states is ensured by the fact that they are in the sub-space of the solutions that is cut out by the non-linearities of the Dirac equations, and this is encouraging, but since the space of the solutions has been halved one may wonder how we may double it in order to recover the correct number of independent components, and this can be done by doubling the Dirac equations allowing both signs of the mass \cite{Fabbri:2012ag}; the necessity to have all four independent components means the possibility to describe antimatter as well as matter, and in the above paper these two fields are described as the two independent solutions of the two complementary Dirac equations differing for the sign of the mass term, although having a negative mass term does not mean that negative-mass states are present. We will discuss this in more detail.

To better explain the situation, we will start recalling the definition of matter/antimatter duality, which we will take here in its most general form: according to such general definition, antimatter is matter with all quantum numbers reversed, where not only helicity but also mass, beside all charges, are inverted: as it has already been discussed in \cite{Fabbri:2012ag}, although this implies that we have two forms of matter field equations differing by the sign of the mass term, this does not mean that masses will become negative but merely that there is a rearrangement of the physical components within the spinor, so that such inversion does not imply that matter has a positive mass and antimatter has a negative mass but simply that matter and antimatter are represented by two spinors with different structures: by calling the matter and antimatter fields respectively as $e$ and $p$ then matter and antimatter were related by the relationship $p\!=\!\boldsymbol{\gamma}e$ as it should be in order to reproduce the known scattering amplitudes, and since the negative-mass spinors were always suppressed when field equations were imposed then only positive-mass spinors were present; additionally, we proved that both matter and antimatter had only positive energies, as it should be for consistency. According to this general definition of the matter/antimatter duality, there is not a single matter field with positive/negative energies but two complementary matter/antimatter fields with only positive energies. But in this situation matter and antimatter are no longer complex conjugate of one another, and instead they are two independent fields, so that the passage between one to the other is not achieved in terms of charge conjugation but simply by changing the sign of the charge 
$q\!\rightarrow\!-q$ only: in particular neutral spinors are not described by real-valued spinors accomplished in terms of self-charge conjugated fields but simply in terms of complex-valued spinors with the supplementary condition $q\!=\!0$ solely; because in this approach neutrality is no longer obtained in terms of constraints on the components of the spinor then neutral fields are no longer described with half degrees of freedom, and thus they are no longer forced to be massless. In this scheme, there is no longer a reason why neutral massive fields should not exist in general: if dark matter is to be described in terms of neutral massive particles, in this scheme there is room for a dark matter candidate. However, the presence of torsionally-induced non-linearities for these dark matter candidates would change their dynamics.

That a cold bath of neutral massive particles may be able to form condensates which might stretch up to galactic scales is a known conjecture \cite{Silverman:2002qx}; but if torsion is not neglected for the dynamics of these particles then torsional effects may be relevant \cite{Tilquin:2011bu}: that these condensates feel such torsionally-induced non-linear potentials can be seen in the non-relativistic approximation, where
\begin{eqnarray}
&\boldsymbol{\nabla}^{2}\psi\!-\!Y^{2}|\overline{\psi}\psi|^{2}\psi
\!-\!2mY\overline{\psi}\psi\psi\!+\!2m(E\!-\!m)\psi\!\approx\!0
\end{eqnarray}
and if through such a condensate a test body will move, the non-linear corrections to the gravitational field are such that the gravitational acceleration $\vec{a}$ is given by
\begin{eqnarray}
&\mathrm{div}\vec{a}\!=\!-\frac{1}{2}Y|\overline{\psi}\psi|^{2}\!-\!\frac{1}{4}m\overline{\psi}\psi
\end{eqnarray}
as it can be easily checked. The high-density behaviour of the condensate is given by the exact solution
\begin{eqnarray}
&\overline{\psi}\psi\!=\!\frac{1}{2Y}\frac{1}{r}
\end{eqnarray}
so that the squared of the tangential velocity is given by 
\begin{eqnarray}
&v^{2}\!=\!\frac{1}{8Y}
\end{eqnarray}
displaying the constant behaviour of the rotation curves for galactic dark matter; in order to fit also the magnitude, it is necessary that the torsional coupling constant be about $10^{8}$ times the Newton constant, and it will be a universal parameter. All these computations and some of their implications have been discussed in reference \cite{Fabbri:2012zc}.

Our discussion has started in terms of a general critics about the problem of the cosmological constant and dark energy moving gradually to tackle the problem of dark matter as well; now we may continue our wandering in the domain of cosmology to face another problem, namely the problem of gravitational singularities, discussing a common misconception: such a problem is based on the Hawking-Penrose theorem, which states that in general gravitationally-produced singular matter distributions are present. We know that the torsionally-induced non-linear potentials are repulsive, so these should counterbalance the gravitational attraction, and it could be that those singularities are avoided. There is a misconception in the formulation of this problem, however.

As it could be demonstrated, the torsion-spin coupling is what contributes the most to gravitational singularities of spinorial matter fields, and so it appears not only that torsion would not solve this problem but also that it would worsen the singularity \cite{k}; the misconception is that we cannot consider only the gravitational field equations in treating the problem. If it is true that the torsionally-induced non-linear interaction increases the amount of energy sourcing gravitation, therefore augmenting the gravitational pull, nevertheless it is also true that the gravitational field equations cannot be the only equations to be taken; we need also to consider the matter field equations, where the torsionally-induced non-linear potentials are repulsive: as the scaling properties of all terms involved are such that on shorter and shorter distances the non-linearities of the gravitational field equation become less and less relevant than the non-linearities of the matter field equations, the repulsion will be more and more relevant than the attraction, and eventually it would become dominant. The presence of the torsion-spin coupling does contribute to both the gravitational attraction and the spinorial intrinsic repulsion, but they are relevant at different scales: as the singularity is about to form the scale would be one at which the gravitational pull would be negligible compared to the intrinsic repulsion of fermionic matter fields. And in fact, as it has been demonstrated in \cite{Fabbri:2011mg}, no matter the gravitational effects the non-linear terms in the matter field equations will always be able to prevent these singularities to form.

From this discussion one point emerges: it is precisely the self-interacting character of the matter field itself what forbids the singularity formation to occur.

Here self-interactions for matter forbid singularity formation in the same way in which usually evading singularities is achieved in terms of quantum effects \cite{b-o-s}.

The fact that self-interacting matter and quantization seems to have analogous effects is what ultimately allows to address some of the open problems in the standard models of cosmology and particle physics in terms of non-linear potentials, from the problem of dark energy with dark matter and avoidance of singularities to the phenomenology of weak interactions of leptons and oscillations of neutrinos; but there is still one issue.

Among all the subjects we have investigated, we have been employing the torsionally-induced non-linear terms in the matter field equations to generate weak-like forces among the electron-neutrino pair, to give rise to oscillation between neutrinos, and to fit the behaviour of flat rotation curves for galaxies in the presence of dark matter, and with torsional coupling constant that had to be tuned to the Fermi constant, to the oscillation length, and to the value of about $10^{8}$ times the Newton constant, respectively; these results are known in the literature, but the problem was that one might have asked how it were possible for these effects to correspond to different values of the coupling constants, and the answer is that in this theory different coupling constants are allowed so long as we consider different fermionic fields because, as already noticed, having included all terms involving torsion resulted into an effective theory in which the original torsion constants were re-expressed in terms of coupling constants of the fermionic fields, as if we could perform on each fermion field an independent renormalization.

The considered problems of the standard models of cosmology and particle physics are altogether addressed in terms of non-linear effects within a single scheme.

This ends our survey on the standard models.
\section{Discussion}
In what we have been doing so far, we have started by considering the completely antisymmetric torsion completion of gravity with electrodynamics for a geometry filled with two Dirac matter fields, assigning the most general interaction between geometry and matter but without mixing terms for the two fermionic matter fields, in an action that was up to terms of dimension four and polynomial in these: of this model, we have derived the conservation laws and conserved quantities, together with all field equations, we have acknowledge that the cosmological constant and the two masses of the fermions could be interpreted as integration constants and therefore allowed as a matter of generality, we have written the field equations in the equivalent form in which all torsional contributions were converted into torsionally-induced self-spinorial interactions of a specific structure and with three coupling constants that had to be taken as undetermined; we have seen how some open problem in physics could be solved by non-linear potentials replacing fermion-field quantization. So dramatic as it may seem that something could replace field quantization, this possibility rises the question about whether field quantization might still be considered to be an essential operation.

Thus for the sake of argumentation we ask: for what reason we quantize fields, by requiring them to be expanded in a basis of creation/annihilation operators and imposing some sort of commutation relationships among their components? This idea, so far as we are aware, had at its core an argument of analogy, that is repeating for fields what has already been found to be successful for point particles: the particle quantization, or first quantization, up-graded the classical mechanics to quantum mechanics, and similarly field quantization, or second-quantization, should bring classical fields into quantum fields; the problem with this argument is that here, there is no real analogy between particles and fields, and even if there were an analogy allowing us to apply the same prescription in both instances, the analogy would only tell us that such quantization may work similarly in the two circumstances, which might mean they are equally good, but also equally bad. Of course, stating that quantization is not a good prescription for fields would imply that quantization is not a good prescription for particles.

Assuming that particle quantization is not an appropriate prescription too, then why does it seem to work so well in giving rise to all the successful predictions of quantum mechanics? The fact is that classical mechanics deals with point particles, which are problematic because they are singular matter distributions: therefore requiring the variables to be operators subject to commutation relations like $[p_{\mu},x^{\nu}]\!=\!i\hbar\delta_{\mu}^{\nu}$ or in position representation as given by $p_{\mu}\!=\!i\hbar\nabla_{\mu}$ means demanding for extended matter distributions, that is the classical fields of quantum mechanics; the reason for which particle quantization, which might be a bad prescription, works so nicely is that it applies to classical mechanics, which is also intrinsically problematic, and the two problems cancel one another, resulting into a quantum mechanics of classical fields that is successful. But there is another question.

If it is true that starting from non-relativistic classical mechanics and performing particle quantization we get the non-relativistic classical fields of quantum mechanics as a lucky accident, then it should be possible to build the quantum mechanics and its non-relativistic classical field content directly. Because quantum mechanics is constructed on the Schr\"{o}dinger non-relativistic classical field equation, this question may be translated into asking if it is possible to get the Schr\"{o}dinger equation starting from first principles: Feynman once said that there is no pattern that made Schr\"{o}dinger guess the equation carrying his name, but today we know that the Schr\"{o}dinger equation, or the Pauli-Schr\"{o}dinger equation, is the non-relativistic weak-field approximation of the Dirac equation, obtainable from first principles, as we did here.

Quantum mechanics is derivable as a non-relativistic weak-field limit from relativistic classical fields.
\section*{Conclusion}
In this paper, we have starter from the most general dynamics involving the completely antisymmetric torsion completion of gravity with electrodynamics for two non-mixing Dirac matter fields: we have derived the effective theory trying to use the resulting non-linear potentials of the matter field equations in order to address some of the open problems in the standard models of cosmology and particle physics; this has been done by exploiting the non-linear potentials instead of fermionic-field quantization.

We have started by discussing the fact that the cosmological constant problem could be solved by merely acknowledging that a cosmological constant may always be allowed to be present and because it can be thought as an integration constant then its value has to be determined empirically; this solution, however, requires that none of the additional contributions are present, and this means that there can be neither a mechanism of mass generation nor any cosmological version of the Casimir effect, or in equivalent terms, there can be neither a breakdown of any given symmetry nor some sort of zero-point energy for the vacuum, respectively: dropping the breakdown of the symmetry means that the asymmetric situation we see is such because it has always been asymmetric, and in this scheme we have to find a way to assign masses to fields that are supposed to be massive, and to recover the phenomenology of the weak forces, the former of these tasks being achieved by acknowledging that mass terms may always be present, while the latter of these tasks being accomplished by the fact that the non-linear potentials are structurally identical to the weak forces thus yielding the same phenomenology; further, dropping the concept of zero-point energy for the vacuum means that we have to recover the phenomenology of the Casimir effect between two plates, and this can be done by having the Casimir force as the result of electrodynamic interactions between the two source fields. About the problem of cold dark matter, it has been acknowledged that a bath of neutral massive particles is allowed whenever we have the possibility to require the vanishing of the charge without forcing the mass to vanish, which can be done if neutral fields could be obtained without the halving of their degrees of freedom: this situation would require that neutral fields are not self-charge conjugated fields, or more in general that antimatter is not matter after the process of charge conjugation, but simply two independent fields, which would have to be defined without Grassmann variables: dropping the assumption of having Grassmann-valued fermions means that we have to reproduce the Fermi-Dirac statistic and to ensure the positivity of the energy for both matter and antimatter fermions, and again these issues were addressed by exploiting the non-linearities of the Dirac matter field equations, as discussed in \cite{Fabbri:2012ag}. Eventually, we have also reported that in \cite{Fabbri:2011mg} it has been demonstrated how this situation prevents gravitationally-produced singularities of the considered Dirac matter field distribution to form.

We have specified that employing the most general least-order derivative torsional completion of gravity with fermionic Dirac matter fields allowed us to have these results altogether described within a single model.

In the present article, we have obtained our results by exploiting the non-linear potentials within the matter field equations; on the other hand, we have used no fermion field-quantization whatsoever: the fact that effects of fermion field-quantization may be recovered by torsionally-induced non-linear potentials in matter field equations points toward a connection between spacetime torsion and field quantization, a connection already noticed in \cite{Fabbri:2013gza} and which we deepened here. As it is clear, this does not mean that such a connection is established, and more has to be done to investigate this link between these two complementary concepts; of course, further papers will be devoted to that. But nevertheless, it was necessary to stress that this parallel is quite interesting.

The parallel between field-quantization of fermions and torsionally-induced non-linear potentials in the matter field equations might look impertinent, but if it turns to be true that effects due to field-quantization prescriptions may be replaced by those coming from self-interacting potentials in the matter field equations this would not be so surprising after all, if we think that the non-linear terms of the matter field equations are a direct consequence of the coupling between matter and its underlying background considered in the most general situation while the field quantization is a protocol that is forced rather arbitrarily in the model; such a protocol has been kept up for long in part because of its successful predictions but in part also because wrong predictions have never been considered as a hint of a possible falsification, but rather they were euphemistically renamed anomalies, and then assumed to be cancelled by treating them in terms of yet further adjustments, some of which are still unknown.

It is clear that such behaviour may hinder the process of searching for a fundamental theory in physics and it may be the time to stop and reverse this tendency.
\appendix \footnotesize
\section{Fierz identities}
\label{a}
In this paper, we have made a considerable use of the Fierz identities every time we had to rearrange the components of a pair of spinor fields or the components of a single spinor field, but we have never mentioned what these identities actually were, and therefore we will take the opportunity in this appendix to list some of these Fierz identities: the first is certainly given by the general identity valid for any pair of spinor fields according to the expression
\begin{eqnarray}
\nonumber
&\psi_{1}\overline{\psi}_{2}\!=\!\frac{1}{4}\overline{\psi}_{2}\psi_{1}\mathbb{I}
\!-\!\frac{1}{2}\overline{\psi}_{2}\boldsymbol{\sigma}_{ij}\psi_{1}\boldsymbol{\sigma}^{ij}
\!-\!\frac{1}{4}i\overline{\psi}_{2}\boldsymbol{\pi}\psi_{1}i\boldsymbol{\pi}+\\
\nonumber
&+\frac{1}{4}\overline{\psi}_{2}\boldsymbol{\gamma}_{i}\psi_{1}\boldsymbol{\gamma}^{i}
\!-\!\frac{1}{4}\overline{\psi}_{2}\boldsymbol{\gamma}_{i}\boldsymbol{\pi}\psi_{1}
\boldsymbol{\gamma}^{i}\boldsymbol{\pi}
\end{eqnarray}
which can be contracted so that one can derive the additional identity determining the rearrangement of two spinor fields as
\begin{eqnarray}
\nonumber
&\overline{\psi}_{1}\boldsymbol{\gamma}_{a}\psi_{1}
\overline{\psi}_{2}\boldsymbol{\gamma}^{a}\psi_{2}
\!=\!\overline{\psi}_{2}\psi_{1}\overline{\psi}_{1}\psi_{2}
\!+\!i\overline{\psi}_{2}\boldsymbol{\pi}\psi_{1}i\overline{\psi}_{1}\boldsymbol{\pi}\psi_{2}-\\
\nonumber
&-\frac{1}{2}\overline{\psi}_{2}\boldsymbol{\gamma}_{i}\psi_{1}
\overline{\psi}_{1}\boldsymbol{\gamma}^{i}\psi_{2}
\!-\!\frac{1}{2}\overline{\psi}_{2}\boldsymbol{\gamma}_{i}\boldsymbol{\pi}\psi_{1}
\overline{\psi}_{1}\boldsymbol{\gamma}^{i}\boldsymbol{\pi}\psi_{2}
\end{eqnarray}
as it is easy to see; the single spinor field would be such that
\begin{eqnarray}
\nonumber
&\psi\overline{\psi}\!=\!\frac{1}{4}\overline{\psi}\psi\mathbb{I}
\!-\!\frac{1}{2}\overline{\psi}\boldsymbol{\sigma}_{ij}\psi\boldsymbol{\sigma}^{ij}
\!-\!\frac{1}{4}i\overline{\psi}\boldsymbol{\pi}\psi i\boldsymbol{\pi}+\\
\nonumber
&+\frac{1}{4}\overline{\psi}\boldsymbol{\gamma}_{i}\psi\boldsymbol{\gamma}^{i}
\!-\!\frac{1}{4}\overline{\psi}\boldsymbol{\gamma}_{i}\boldsymbol{\pi}\psi
\boldsymbol{\gamma}^{i}\boldsymbol{\pi}
\end{eqnarray}
which can be contracted so that one can derive the additional 
\begin{eqnarray}
\nonumber
&\overline{\psi}\boldsymbol{\gamma}_{a}\psi\overline{\psi}\boldsymbol{\gamma}^{a}\psi
\!=\!-\overline{\psi}\boldsymbol{\gamma}_{a}\boldsymbol{\pi}\psi
\overline{\psi}\boldsymbol{\gamma}^{a}\boldsymbol{\pi}\psi=\\
\nonumber
&=|\overline{\psi}\psi|^{2}\!+\!|i\overline{\psi}\boldsymbol{\pi}\psi|^{2}
\label{id1}\\
\nonumber
&\overline{\psi}\boldsymbol{\gamma}_{a}\boldsymbol{\pi}\psi
\overline{\psi}\boldsymbol{\gamma}^{a}\psi=0
\label{id2}
\end{eqnarray}
together with the complementary
\begin{eqnarray}
\nonumber
&4i\overline{\psi}\boldsymbol{\sigma}_{ak}\psi i\overline{\psi}\boldsymbol{\sigma}^{bk}\psi
\!-\!|\overline{\psi}\psi|^{2}\delta_{a}^{b}\!=\!
4\overline{\psi}\boldsymbol{\sigma}_{ak}\boldsymbol{\pi}\psi
\overline{\psi}\boldsymbol{\sigma}^{bk}\boldsymbol{\pi}\psi
\!-\!|i\overline{\psi}\boldsymbol{\pi}\psi|^{2}\delta_{a}^{b}=\\
\nonumber
&=\overline{\psi}\boldsymbol{\gamma}_{a}\boldsymbol{\pi}\psi
\overline{\psi}\boldsymbol{\gamma}^{b}\boldsymbol{\pi}\psi
\!-\!\overline{\psi}\boldsymbol{\gamma}_{a}\psi
\overline{\psi}\boldsymbol{\gamma}^{b}\psi\\
\nonumber
&4i\overline{\psi}\boldsymbol{\sigma}_{ak}\psi
\overline{\psi}\boldsymbol{\sigma}^{bk}\boldsymbol{\pi}\psi
=-i\overline{\psi}\boldsymbol{\pi}\psi\overline{\psi}\psi\delta_{a}^{b}
\end{eqnarray}
and also the following
\begin{eqnarray}
\nonumber
&2i\overline{\psi}\boldsymbol{\sigma}_{ik}\psi
\overline{\psi}\boldsymbol{\gamma}^{i}\psi
=i\overline{\psi}\boldsymbol{\pi}\psi\overline{\psi}\boldsymbol{\gamma}_{k}\boldsymbol{\pi}\psi\\
\nonumber
&2\overline{\psi}\boldsymbol{\pi}\boldsymbol{\sigma}_{ik}\psi
\overline{\psi}\boldsymbol{\gamma}^{i}\psi
=\overline{\psi}\psi\overline{\psi}\boldsymbol{\gamma}_{k}\boldsymbol{\pi}\psi\\
\nonumber
&2i\overline{\psi}\boldsymbol{\sigma}_{ik}\psi
\overline{\psi}\boldsymbol{\gamma}^{i}\boldsymbol{\pi}\psi
=i\overline{\psi}\boldsymbol{\pi}\psi\overline{\psi}\boldsymbol{\gamma}_{k}\psi\\
\nonumber
&2\overline{\psi}\boldsymbol{\pi}\boldsymbol{\sigma}_{ik}\psi
\overline{\psi}\boldsymbol{\gamma}^{i}\boldsymbol{\pi}\psi
=\overline{\psi}\psi\overline{\psi}\boldsymbol{\gamma}_{k}\psi
\end{eqnarray}
together with
\begin{eqnarray}
\nonumber
&\overline{\psi}\psi i\overline{\psi}\boldsymbol{\sigma}_{ab}\psi\!-\!
i\overline{\psi}\boldsymbol{\pi}\psi\overline{\psi}\boldsymbol{\sigma}_{ab}\boldsymbol{\pi}\psi
\!=\!\frac{1}{2}\overline{\psi}\boldsymbol{\gamma}^{j}\psi
\overline{\psi}\boldsymbol{\gamma}^{k}\boldsymbol{\pi}\psi\varepsilon_{jkab}
\end{eqnarray}
and as it is clear, many more of such identities actually exist but for purpose of the present paper these are enough.

\end{document}